\documentclass[fleqn]{article}
\usepackage{espcrc2}
\newbox\dummybox\setbox\dummybox=\hbox{}
{\setbox0=\hbox{{\vrule width0pt} ({hep-lat/0000000})}\wd\dummybox=\wd0}
\newcommand\noeprint[1]{\hbox{#1\copy\dummybox}}
\begin{document}
\onecolumn

\vskip 8mm
{\raggedright \Large Contents}
\vskip 8mm

Preface \dotfill v

Committees \dotfill vii

\section*{PLENARY PRESENTATIONS} 

Unquenched QCD simulation results

S. Aoki \dotfill \hbox{3 (hep-lat/0011074)}

Lattice QCD at finite temperature

S. Ejiri \dotfill \hbox{19 (hep-lat/0011006)}

QCD vacuum structure

M. Garc\'{\i}a P\'erez \dotfill \hbox{27 (hep-lat/0011026)}

Recent developments in superstring theory

A. Sen \dotfill \hbox{35 (hep-lat/0011073)}

Static correlation lengths in QCD at high temperature and finite densities

O. Philipsen \dotfill \hbox{49 (hep-lat/0011019)}

Non-equilibrium field theory

D. B\"odeker \dotfill \hbox{61 (hep-lat/0011077)}

QCD at a finite density of static quarks

S. Chandrasekharan \dotfill \hbox{71 (hep-lat/0011022)}

Non-perturbative renormalization in lattice field theory

S. Sint \dotfill \hbox{79 (hep-lat/0011081)}

Large-N matrix models and applications

H. Kawai \dotfill \noeprint{95}

Discrete Lorentzian quantum gravity

R. Loll \dotfill \hbox{96 (hep-th/0011194)}

Non-lattice determinations of the light quark masses

H. Leutwyler \dotfill \hbox{108 (hep-ph/0011049)}

Quark masses on the lattice: Light and heavy

V. Lubicz \dotfill \hbox{116 (hep-lat/0012003)}

Nonrelativistic bound states in quantum field theory

A.V. Manohar and I.W. Stewart \dotfill \hbox{130 (hep-lat/0012002)}

Light hadron weak matrix elements

L. Lellouch \dotfill \hbox{142 (hep-lat/0011088)}

Heavy quark physics on the lattice

C. Bernard \dotfill \hbox{159 (hep-lat/0011064)}

Domain wall fermions and applications

P.M. Vranas \dotfill \hbox{177 (hep-lat/0011066)}

Lattice chiral gauge theories

M. Golterman \dotfill \hbox{189 (hep-lat/0011027)}

The search for the QGP: A critical appraisal

H. Satz \dotfill \hbox{204 (hep-ph/0009099)}

Lattice gauge theory: A retrospective

M. Creutz \dotfill \hbox{219 (hep-lat/0010047)}

\section*{PARALLEL PRESENTATIONS} 

\subsection*{A. Spectrum and quark masses} 

Full QCD light hadron spectrum and quark masses: Final results from CP-PACS

A. Ali Khan, S. Aoki, G. Boyd, R. Burkhalter, S. Ejiri, M. Fukugita, S. Hashimoto, N. Ishizuka,

Y. Iwasaki, K. Kanaya, T. Kaneko, Y. Kuramashi, T. Manke, K.-I. Nagai, M. Okawa, H.P. Shanahan,

A. Ukawa and T. Yoshi\'e (CP-PACS collaboration) \dotfill \hbox{229 (hep-lat/0010078)}

Light hadron spectroscopy with two flavors of $O(a)$ improved dynamical quarks

S. Aoki, R. Burkhalter, M. Fukugita, S. Hashimoto, K.-I. Ishikawa, N. Ishizuka, Y. Iwasaki,

K. Kanaya, T. Kaneko, Y. Kuramashi, M. Okawa, T. Onogi, S. Tominaga, N. Tsutsui,
A. Ukawa,

N. Yamada and T. Yoshi\'e (JLQCD collaboration) \dotfill \hbox{233 (hep-lat/0010086)}

Quark loop effects with an improved staggered fermion action

C. Bernard, T. Burch, T.A. DeGrand, C.E. DeTar, S. Gottlieb, U.M. Heller, K. Orginos, R.L. Sugar

and D. Toussaint \dotfill \hbox{237 (hep-lat/0010065)}

Effects of non-perturbatively improved dynamical fermions in UKQCD simulations

A.C. Irving (UKQCD collaboration) \dotfill \hbox{242 (hep-lat/0010012)}

Baryon masses in the $1/N$ expansion

E. Jenkins \dotfill \hbox{246 (hep-lat/0011044)}

$N^*$ mass spectrum from and anisotropic action

F.X. Lee \dotfill \hbox{251 (hep-lat/0011060)}

Scalar and tensor glueballs on asymmetric coarse lattices

C. Liu \dotfill \hbox{255 (hep-lat/0010007)}

A derivation of Regge trajectories in large$-N$ transverse lattice QCD

A.D. Patel \dotfill \hbox{260 (hep-lat/0012004)}

Light quark masses for dynamical, non-perturbatively $O(a)$ improved Wilson fermions

D. Pleiter (QCDSF and UKQCD collaborations) \dotfill \hbox{265 (hep-lat/0010063)}

$N^*$ spectrum using an $O(a)$ improved fermion action

D.G. Richards (LHPC and UKQCD collaborations) \dotfill \hbox{269 (hep-lat/0011025)}

Hadron masses from dynamical, non-perturbatively $O(a)$ improved Wilson fermions

H. St\"uben (QCDSF and UKQCD collaborations) \dotfill \hbox{273 (hep-lat/0011045)}

Light hadronic physics using domain wall fermions in quenched lattice QCD

M. Wingate \dotfill \hbox{277 (hep-lat/0009023)}

\subsection*{B. Hadronic matrix elements} 

Calculation of $K\rightarrow\pi$ matrix elements in quenched domain-wall QCD

A. Ali Khan, S. Aoki, Y. Aoki, R. Burkhalter, S. Ejiri, M. Fukugita, S. Hashimoto, N. Ishizuka,

Y. Iwasaki, T. Izubuchi, K. Kanaya, T. Kaneko, Y. Kuramashi, K.-I. Nagai, J. Noaki, M. Okawa,

H.P. Shanahan, Y. Taniguchi, A. Ukawa and T. Yoshi\'e (CP-PACS collaboration) \dotfill \hbox{283 (hep-lat/0011007)}

Kaon B-parameter from quenched domain-wall QCD

A. Ali Khan, S. Aoki, Y. Aoki, R. Burkhalter, S. Ejiri, M. Fukugita, S. Hashimoto, N. Ishizuka,

Y. Iwasaki, T. Izubuchi, K. Kanaya, T. Kaneko, Y. Kuramashi, K.-I. Nagai, J. Noaki, M. Okawa,

H.P. Shanahan, Y. Taniguchi, A. Ukawa and T. Yoshi\'e (CP-PACS collaboration) \dotfill \hbox{287 (hep-lat/0010079)}

$K\rightarrow\pi\pi$ with domain wall fermions: Lattice matrix elements

T. Blum (RBC collaboration) \dotfill \hbox{291 (hep-lat/0011042)}

Domain wall fermion calculation of nucleon $g_A/g_V$

T. Blum, S. Ohta and S. Sasaki \dotfill \hbox{295 (hep-lat/0011011)}

\newpage

Four-quark operators in hadrons

S. Capitani, M. G\"ockeler, R. Horsley, B. Klaus, W. K\"urzinger, D. Petters, D. Pleiter, P.E.L. Rakow,

S. Schaefer, A. Sch\"afer and G. Schierholz \dotfill \hbox{299 (hep-lat/0010043)}

Moments of structure functions in full QCD

D. Dolgov, R. Brower, S. Capitani, J.W. Negele, A. Pochinsky, D. Renner, N. Eicker, T. Lippert,

K. Schilling, R.G. Edwards and U.M. Heller \dotfill \hbox{303 (hep-lat/0011010)}

Determining $g_A$ using non-perturbatively $O(a)$ improved Wilson fermions

R. Horsley (QCDSF and UKQCD collborations) \dotfill \hbox{307 (hep-lat/0010059)}

Strange magnetic moment of the nucleon from lattice QCD

N. Mathur and S.-J. Dong \dotfill \hbox{311 (hep-lat/0011015)}

$K\rightarrow\pi\pi$ decays with domain wall fermions: Towards physical results

R.D. Mawhinney (RBC collaboration) \dotfill \hbox{315 (hep-lat/0010030)}

Disconnected electromagnetic form factors

W. Wilcox \dotfill \hbox{319 (hep-lat/0010060)}

\vspace{-0.7\baselineskip}
\subsection*{C. Heavy quark physics} 

Quenched charmonium spectrum on anisotropic lattices

A. Ali Khan, S. Aoki, R. Burkhalter, S. Ejiri, M. Fukugita, S. Hashimoto, N. Ishizuka, Y. Iwasaki,

K. Kanaya, T. Kaneko, Y. Kuramashi, K.-I. Nagai, M. Okamoto, M. Okawa, H.P. Shanahan,

Y. Taniguchi, A. Ukawa and T. Yoshi\'e (CP-PACS collaboration) \dotfill \hbox{325 (hep-lat/0011005)}

Differential decay rate for $B\rightarrow\pi l\nu$ semileptonic decays

S. Aoki, R. Burkhalter, M. Fukugita, S. Hashimoto, K.-I. Ishikawa, N. Ishizuka, Y. Iwasaki,

K. Kanaya, T. Kaneko, Y. Kuramashi, M. Okawa, T. Onogi, S. Tominaga, N. Tsutsui,
A. Ukawa,

N. Yamada and T. Yoshi\'e (JLQCD collaboration) \dotfill \hbox{329 (hep-lat/0011008)}

On the short distance nonperturbative corrections in heavy quark expansion

S. Arunagiri \dotfill \hbox{333 (hep-lat/0011061)}

Heavy quark phenomenology from lattice QCD

D. Becirevic \dotfill \hbox{337 (hep-lat/0011075)}

Relativistic quarkonia from anisotropic lattices

P. Chen, X. Liao and T. Manke \dotfill \hbox{342 (hep-lat/0010069)}

$f_B$ for various actions: Approaching the continuum limit with dynamical fermions

C. Bernard, S. Datta, C. DeTar, S. Gottlieb, U.M. Heller, J. Hetrick, C. McNeile, K. Orginos,

R. Sugar and D. Toussaint \dotfill \hbox{346 (hep-lat/0011029)}

Quenched and first unquenched lattice HQET determination of the $B_s-$meson width difference

V. Gim\`enez and J. Reyes \dotfill \hbox{350 (hep-lat/0010048)}

Perturbative renormalization of the $\Delta B=2$ four-quark operators in lattice NRQCD

S. Hashimoto, K.-I. Ishikawa, T. Onogi, M. Sakamoto, N. Tsutsui and\par N. Yamada \dotfill \hbox{354 (hep-lat/0010056)}

Heavy-light meson spectrum with and without NRQCD

R. Lewis and R.M. Woloshyn \dotfill \hbox{359 (hep-lat/0010001)}

The unquenched $\Upsilon$ spectrum

L. Marcantonio, P. Boyle, C.T.H. Davies, J. Hein and \par J. Shigemitsu (UKQCD collaboration) \dotfill \hbox{363 (hep-lat/0011053)}

Leptonic and semi-leptonic B decays

C.M. Maynard (UKQCD collaboration) \dotfill \hbox{367 (hep-lat/0010016)}

Can one study heavy meson semileptonic decays on coarse anisotropic lattices?

J. Shigemitsu \dotfill \hbox{371 (hep-lat/0010029)}

Unquenched charmonium with NRQCD

C. Stewart and R. Koniuk \dotfill \hbox{375 (hep-lat/0010015)}

\newpage

$B^0-\bar{B}^0$ mixing with quenched lattice NRQCD

N. Yamada, S. Aoki, R. Burkhalter, M. Fukugita, S. Hashimoto, K.-I. Ishikawa, N. Ishizuka,

Y. Iwasaki, K. Kanaya, T. Kaneko, Y. Kuramashi, M. Okawa, T. Onogi, S. Tominaga, N. Tsutsui,

A. Ukawa and T. Yoshi\'e (JLQCD collaboration) \dotfill \hbox{379 (hep-lat/0010089)}

\subsection*{D. Finite temperature} 

Screeing in hot $SU(2)$ gauge theory and propagators in 3d adjoint Higgs model

A. Cucchieri, F. Karsch and P. Petreczky \dotfill \hbox{385 (hep-lat/0010023)}

Deconfinement through chiral transition in 2 flavour QCD

S. Digal, E. Laermann and H. Satz \dotfill \hbox{389 (hep-lat/0010046)}

Thermodynamics of free domain wall and overlap fermions

G.T. Fleming \dotfill \hbox{393 (hep-lat/0011069)}

Percolation and critical behaviour in $SU(2)$ gauge theory

S. Fortunato, F. Karsch, P. Petreczky and H. Satz \dotfill \hbox{398 (hep-lat/0010026)}

Dirac eigenvalues and eigenvectors at finite temperature

M. G\"ockeler, H. Hehl, P.E.L. Rakow, A. Sch\"afer, W. S\"oldner and T. Wettig \dotfill \hbox{402 (hep-lat/0010049)}

$SU(2)$ lattice gauge theory at non-zero temperature with fixed holonomy boundary condition

E.-M. Ilgenfritz, B. Martenmyanov, M. M\"uller-Preussker and A.I. Veselov \dotfill \hbox{407 (hep-lat/0011051)}

Flavor and quark mass dependence of QCD thermodynamics

F. Karsch, E. Laermann, A. Peikert, Ch. Schmidt and S. Stickan \dotfill \hbox{411 (hep-lat/0010040)}

Obstructions to dimensional reduction in hot QCD

S. Gupta \dotfill \hbox{415 (hep-lat/0010032)}

Spinodal decomposition in finite temperature $SU(2)$ and $SU(3)$

T.R. Miller and M.C. Ogilvie \dotfill \hbox{419 (hep-lat/0010055)}

Towards the application of the maximum entropy method to finite temperature Upsilon spectroscopy

M. Oevers, C. Davies and J. Shigemitsu \dotfill \hbox{423 (hep-lat/0009031)}

Damping and the Hartree ensemble approximation

M. Sall\'e, J. Smit and J.C. Vink \dotfill \hbox{427 (hep-lat/0010054)}

Finite temperature simulations from quantum field dynamics?

M. Sall\'e, J. Smit and J.C. Vink \dotfill \hbox{431 (hep-lat/0010062)}

Charmonium in finite temperature lattice QCD

T. Umeda, R. Katayama, H. Matsufuru and O. Miyamura \dotfill \hbox{435 (hep-lat/0010090)}

\subsection*{E. Finite density} 

Topology in full QCD with 2 colours at finite temperature and density

B. All\'es, M. D'Elia, M.-P. Lombardo and M. Pepe \dotfill \hbox{441 (hep-lat/0010068)}

Dirac and Gor'kov spectra in two color QCD with chemical potential

E. Bittner, M.-P. Lombardo, H. Markum and R. Pullirsch \dotfill \hbox{445 (hep-lat/0010018)}

Flop transitions in cuprate and color superconductors: From $SO(5)$ to $SO(10)$ unification?

S. Chandrasekharan, V. Chudnovsky, B. Schlittgen and U.-J. Wiese \dotfill \hbox{449 (hep-lat/0011054)}

Chemical potential response of meson masses at finite temperature

Ph. de Forcrand, T. Hashimoto, S. Hioki, Y. Liu, H. Matsufuru, O. Miyamura, A. Nakamura,

T. Takaishi and T. Umeda (QCD-TARO collaboration) \dotfill \hbox{453 (hep-lat/0011013)}

Two-colour QCD at finite fundamental quark-number density and related theories

S.J. Hands, J.B. Kogut, S.E. Morrison and D.K. Sinclair \dotfill \hbox{457 (hep-lat/0010028)}

Numerical study of dense adjoint 2-color matter

S. Hands, I. Montvay, M. Oevers, L. Scorzato and J. Skullerud \dotfill \hbox{461 (hep-lat/0010085)}

\newpage

Monte Carlo study of two-color QCD with finite chemical potential:

Status report of Wilson fermion simulation

S. Muroya, A. Nakamura and C. Nonaka \dotfill \hbox{469 (hep-lat/0010073)}

\subsection*{F. Topology and vacuum} 

Vortices in $SO(3) \times Z(2)$ simulations

A. Alexandru and R.W. Haymaker \dotfill \hbox{475 (hep-lat/0009012)}

The simple center projection of $SU(2)$ gauge theory

B.L.G. Bakker, A.I. Veselov and M.A. Zubkov \dotfill \hbox{478 (hep-lat/0007022)}

Center vortices in MCG: Finite-size and gauge-copy effects

R. Bertle, M. Faber, J. Greensite and \v{S}. Olejn\'{\i}k \dotfill \hbox{482 (hep-lat/0010058)}

Magnetic condensation and confinement in lattice gauge theory

P. Cea and L. Cosmai \dotfill \hbox{486 (hep-lat/0010034)}

Vortices in $SU(2)$ lattice gauge theory

S. Cheluvaraja \dotfill \hbox{490 (hep-lat/0010042)}

The interaction between center monopoles in $SU(2)$ Yang-Mills

Ph. de Forcrand, M. D'Elia and M. Pepe \dotfill \hbox{494 (hep-lat/0010072)}

Laplacian gauge and instantons

Ph. de Forcrand and M. Pepe \dotfill \hbox{498 (hep-lat/0010093)}

Vortices versus monopoles in color confinement

L. Del Debbio, A. Di Giacomo and B. Lucini \dotfill \hbox{502 (hep-lat/0010044)}

Study of Leutwyler-Smilga regimes: Lessons for full QCD simulations

S. D\"urr \dotfill \hbox{506 (hep-lat/0010037)}

Center vortices at strong couplings

M. Faber, J. Greensite and \v{S}. Olejn\'{\i}k \dotfill \hbox{510 (hep-lat/0010077)}

A smoother approach to scaling by suppressing monopoles and vortices

R.V. Gavai \dotfill \hbox{514 (hep-lat/0010022)}

Vortex structure of the vacuum and confinement

T.G. Kov\'acs and E.T. Tomboulis \dotfill \hbox{518 (hep-lat/0010076)}

Topologically non-trivial configurations in 3-dimensional Yang-Mills theory

P. Majumdar and D.-S. Shin \dotfill \hbox{522 (hep-lat/0010087)}

Gauge independence of Abelian and monopole dominance

F. Shoji \dotfill \hbox{525 (hep-lat/0010088)}

The Gribov ambiguity for maximal Abelian and center gauges in $SU(2)$ lattice gauge theory

J.D. Stack and W.W. Tucker \dotfill \hbox{529 (hep-lat/0011034)}

QCD topology using scale controlled cooling: Densities and cooling invariant observables

I.-O. Stamatescu and C. Weiss \dotfill \hbox{532 (hep-lat/0010075)}

Monopoles, vortices and confinement in $SU(3)$ lattice gauge theory

R. Wensley and J. Stack \dotfill \hbox{537 (hep-lat/0011020)}

\subsection*{G. Confinement and strings} 

Vortices and monopole distributions in $Z(2) \times SO(3)$ lattice gauge theory

A. Alexandru and R.W. Haymaker \dotfill \hbox{543 (hep-lat/0009013)}

Zero temperature string breaking with staggered quarks

C. Bernard, T. Burch, T.A. DeGrand, C.E. DeTar, S. Gottlieb, U.M. Heller, P. Lacock, K. Orginos,

R.L. Sugar and D. Toussaint \dotfill \hbox{546 (hep-lat/0010066)}

Some universal properties of the string breaking

F. Gliozzi \dotfill \hbox{550 (hep-lat/0010084)}

Static three quark potential in the quenched lattice QCD

H. Matsufuru, Y. Nemoto, H. Suganuma, T.T. Takahashi and T. Umeda \dotfill \hbox{554 (hep-lat/0010084)}

Numerical studies of confinement in the lattice Landau gauge

H. Nakajima, S. Furui and A. Yamaguchi \dotfill \hbox{558 (hep-lat/0010083)}

Confining configurations in QCD and relation to rigid strings

R. Parthasarathy \dotfill \hbox{562 (hep-lat/0010017)}

\subsection*{H. Perturbation theory} 

Fermionic loops in numerical stochastic perturbation theory

F. Di Renzo and L. Scorzato \dotfill \hbox{567 (hep-lat/0010064)}

The vacuum polarization: Power corrections beyond OPE?

M. G\"ockeler, R. Horsley, W. K\"urzinger, V. Linke, D. Pleiter, P.E.L. Rakow and \par G. Schierholz \dotfill \hbox{571 (hep-lat/0012010)}

Asymptotically free theories based on discrete subgroups

P. Hasenfratz and F. Niedermayer \dotfill \hbox{575 (hep-lat/0011041)}

Scale setting for $\alpha_s$ beyond leading order

K. Hornbostel, G.P. Lepage and C. Morningstar \dotfill \hbox{579 (hep-lat/0011049)}

Two-loop perturbative quark mass renormalization from large $\beta$ Monte Carlo

K.J. Juge \dotfill \hbox{584 (hep-lat/0011021)}

Nontrivial fixed point in nonabelian models

A. Patrascioiu and E. Seiler \dotfill \hbox{588 (hep-lat/0010052)}

\subsection*{I. Improvement and renormalisation} 

Renormalization constants using quark states in fixed gauge

T. Bhattacharya, R. Gupta and W. Lee \dotfill \noeprint{595}

Improvement and renormalization constants in $O(a)$ improved lattice QCD

T. Bhattacharya, R. Gupta, W. Lee and S. Sharpe \dotfill \hbox{599 (hep-lat/0101007)}

Constructing improved overlap fermions in QCD

W. Bietenholz, N. Eicker, I. Hip and K. Schilling \dotfill \hbox{603 (hep-lat/0011012)}

A nonperturbative determination of $C_A$

S. Collins and C.T.H. Davies (UKQCD collaboration) \dotfill \hbox{608 (hep-lat/0010045)}

Non-perturbative renormalisation with domain wall fermions

C. Dawson (RBC collaboration) \dotfill \hbox{613 (hep-lat/0011036)}

Non-perturbative scaling tests of twisted mass QCD

M. Della Morte, R. Frezzotti, J. Heitger and S. Sint \dotfill \hbox{617 (hep-lat/0010091)}

The perfect Laplace operator for non-trivial boundaries

S. Hauswirth \dotfill \hbox{622 (hep-lat/0010033)}

Progress using generalized lattice Dirac operators to parametrize the fixed-point QCD action

P. Hasenfratz, S. Hauswirth, K. Holland, T. J\"org, F. Niedermayer and U. Wenger \dotfill \hbox{627 (hep-lat/0010061)}

Fixed point $SU(3)$ gauge actions: Scaling properties and glueballs

F. Niedermayer, P. R\"ufenacht and U. Wenger \dotfill \hbox{636 (hep-lat/0011041)}

Perfect gauge actions on anisotropic lattices

P. R\"ufenacht and U. Wenger \dotfill \hbox{640 (hep-lat/0010057)}

Domain wall fermion study of scaling in non-perturbative renormalization of quark composite operators

Y. Zhestkov \dotfill \hbox{644 (hep-lat/0011002)}

\subsection*{J. Topics in gauge theories} 

Finite size scaling analysis of compact QED

G. Arnold, T. Lippert, K. Schilling and T. Neuhaus \dotfill \hbox{651 (hep-lat/0011058)}

\newpage

On the doubling phenomenon in lattice Chern-Simons theories

F. Berruto, M.C. Diamantini and P. Sodano \dotfill \hbox{657 (hep-lat/0011052)}

Zero-momentum modes and chiral limit in compact lattice QED

I.L. Bogolubsky, V.K. Mitrjushkin, M. M\"uller-Preussker and N.V. Zverev \dotfill \hbox{661 (hep-lat/0011024)}

Quasi-local update algorithms for numerical simulations of $d=3$ $SU(2)$ lattice gauge theory in the

dual formulation

N.D. Hari Dass \dotfill \hbox{665 (hep-lat/0011047)}

Current status of the numerical simulations of $d=3$ $SU(2)$ lattice gauge theory in the dual formulation

N.D. Hari Dass and D.-S. Shin \dotfill \hbox{670 (hep-lat/0011038)}

Spinons and holons on the lattice

J. Jersak \dotfill \hbox{675 (hep-lat/0010013)}

\subsection*{K. Gravity and matrix models} 

Spectra of massive QCD Dirac operators from random matrix theory:

All three chiral symmetry breaking patterns

G. Akemann and E. Kanzieper \dotfill \hbox{681 (hep-lat/0010092)}

Simulating simplified versions of the IKKT matrix model

J. Ambj\o rn, K.N. Anagnostopoulos, W. Bietenholz, T. Hotta and J. Nishimura \dotfill \hbox{685 (hep-lat/0009030)}

Computer simulations of 3-d Lorentzian quantum gravity

J. Ambj\o rn, J. Jurkiewicz and R. Loll \dotfill \hbox{689 (hep-lat/0011055)}

High temperature limit of the $N=2$ II-A matrix model

S. Bal and B. Sathiapalan \dotfill \hbox{693 (hep-lat/0011039)}

Matrix model formulation of four dimensional gravity

R. De Pietri \dotfill \hbox{697 (hep-lat/0011033)}

Geometry of 4-d simplicial quantum gravity with a $U(1)$ gauge field

H.S. Egawa, S. Horata and T. Yukawa \dotfill \hbox{701 (hep-lat/0010050)}

Dynamical Regge calculus as lattice gravity

H. Hagura \dotfill \noeprint{704}

Numerical analysis of the double scaling limit in the bosonic part of the II-B matrix model

S. Horata and H.S. Egawa \dotfill \hbox{708 (hep-lat/0010051)}

Towards the lattice study of M-theory

R.A. Janik and J. Wosiek \dotfill \hbox{711 (hep-lat/0011031)}

Reformulating Yang-Mills theory in terms of local gauge invariant variables

P. Majumdar and H.S. Sharatchandra \dotfill \hbox{715 (hep-lat/0010067)}

Maxwell theory from matrix model

H. Takata \dotfill \hbox{718 (hep-lat/0010081)}

\subsection*{L. Chiral fermions} 

Eigenvalues of the hermitian Wilson-Dirac operator and chiral properties of the domain-wall fermion

A. Ali Khan, S. Aoki, Y. Aoki, R. Burkhalter, S. Ejiri, M. Fukugita, S. Hashimoto, N. Ishizuka,

Y. Iwasaki, T. Izubuchi, K. Kanaya, T. Kaneko, Y. Kuramashi, T. Manke, K.-I. Nagai, J. Noaki,

M. Okawa, H.P. Shanahan, Y. Taniguchi, A. Ukawa and \par T. Yoshi\'e (CP-PACS collaboration) \dotfill \hbox{725 (hep-lat/0011032)}

Lattice QCD and chiral mesons

S. Caracciolo, F. Palumbo and R. Scimia \dotfill \hbox{729 (hep-lat/9910039)}

Ginsparg-Wilson relation with $R=(a\gamma_5D)^{2k}$

T.-W. Chiu \dotfill \hbox{733 (hep-lat/0010070)}

\newpage

Exact chiral symmetry for domain wall fermions with finite $L_s$

R.G. Edwards and U.M. Heller \dotfill \hbox{737 (hep-lat/0010035)}

Exact results and approximation schemes for the Schwinger model with the overlap Dirac operator

L. Giusti, C. Hoelbling and C. Rebbi \dotfill \hbox{741 (hep-lat/0011014)}

Residual chiral symmetry breaking in domain-wall fermions

C. Jung, R.G. Edwards, X. Ji and V. Gadiyak \dotfill \hbox{748 (hep-lat/0010094)}

Overlap fermions on a $20^4$ lattice

K.-F. Liu, S.-J. Dong, F.X. Lee and J.B. Zhang \dotfill \hbox{752 (hep-lat/0011072)}

Chiral symmetry in lattice QCD

A.A. Slavnov \dotfill \hbox{756 (hep-lat/0010074)}

Chiral symmetry breaking for domain wall fermions in quenched lattice QCD

L. Wu \dotfill \hbox{759 (hep-lat/0010098)}

\subsection*{M. Chiral gauge theories} 

$U(1)$ chiral gauge theory on lattice with gauge-fixed domain wall fermions

S. Basak and A.K. De \dotfill \hbox{765 (hep-lat/0011018)}

Is the chiral $U(1)$ theory trivial?

V. Bornyakov, A. Hoferichter and G. Schierholz \dotfill \hbox{773 (hep-lat/0011063)}

Theories with global gauge anomalies on the lattice

P. Mitra \dotfill \hbox{777 (hep-lat/0010003)}

The standard model and parity conservation

S.-S. Xue \dotfill \hbox{781 (hep-lat/0010031)}

\subsection*{N. Supersymmetry} 

SUSY Ward identities in $N=1$ SYM theory on the lattice

F. Farchioni, A. Feo, T. Galla, C. Gebert, R. Kirchner, I. Montvay, G. M\"unster and A. Vladikas

(DESY-M\"unster-Roma collaboration) \dotfill \hbox{787 (hep-lat/0010053)}

On the 1-loop lattice perturbation theory of the supersymmetric Ward identities

F. Farchioni, A. Feo, T. Galla, C. Gebert, R. Kirchner, I. Montvay and G. M\"unster

(DESY-M\"unster collaboration) \dotfill \hbox{791 (hep-lat/0011030)}

Supersymmetry on lattice using Ginsparg-Wilson relation

H. So and N. Ukita \dotfill \hbox{795 (hep-lat/0011050)}

\subsection*{O. Algorithms} 

Noisy Monte Carlo algorithm

T. Bakeyev and Ph. de Forcrand \dotfill \hbox{801 (hep-lat/0010006)}

Results on the gluon propagator in lattice covariant gauges

L. Giusti, M.L. Paciello, S. Petrarca, C. Rebbi and B. Taglienti \dotfill \hbox{805 (hep-lat/0010080)}

The role of Monte Carlo within a diagonalization/Monte Carlo scheme

D. Lee \dotfill \hbox{809 (hep-lat/0010096)}

Pad\'e$-Z_2$ stochastic estimator of determinants applied to quark loop expansion of lattice QCD

J.F. Markham and T.D. Kieu \dotfill \hbox{813 (hep-lat/0011001)}

Simulation of $n_f=3$ QCD by hybrid Monte Carlo

T. Takaishi and Ph. de Forcrand \dotfill \hbox{818 (hep-lat/0011003)}

\subsection*{P. Computers} 

QCDOC: A 10-teraflops scale computer for lattice QCD

D. Chen, N.H. Christ, C. Cristian, Z. Dong, A. Gara, K. Garg, B. Joo, C. Kim, L. Levkova, X. Liao,

R.D. Mawhinney, S. Ohta and T. Wettig \dotfill \hbox{825 (hep-lat/0011004)}

Comparing clusters and supercomputers for lattice QCD

S. Gottlieb \dotfill \hbox{833 (hep-lat/0011071)}

Benchmarking MILC code with OpenMP and MPI

S. Gottlieb and S. Tamhankar \dotfill \hbox{841 (hep-lat/0011037)}

The status of APE projects

The APE collaboration: R. Alfieri, R. Di Renzo, E. Onofri, A. Bartoloni, C. Battista, N. Cabbibo,

M. Cosimi, A. Lonardo, A. Michelotti, B. Proietti, F. Rapuano, D. Rossetti, G. Sacco, S. Tassa,

M. Torelli, P. Vicini, Ph. Boucaud, O. P\`ene, W. Errico, G. Magazz, L. Sartori, F. Schifano,

R. Tripiccione, P. De Riso, R. Petronzio, C. Destri, R. Frezzotti, G. Marchesini, U. Gensch,

K. Jansen, A. Kretzschmann, H. Leich, N. Paschendag, D. Pleiter, U. Schwendicke, H. Simma,

R. Sommer, K. Sulanke, P. Wegner, A. Fucci, B. Martin, J. Pech, E. Panizzi, \par and A. Petricola \dotfill \noeprint{846}

Panel discussion: Innovative approaches to high performance computing

G. Bhanot, S. Gottlieb, R. Gupta, M. Okawa, F. Rapuano and R. Mawhinney \dotfill \noeprint{854}

\subsection*{Q. Spin models} 

Critical exponents and equation of state of three-dimensional spin models

M. Campostrini, M. Hasenbusch, A. Pelissetto, P. Rossi and E. Vicari \dotfill \hbox{857 (hep-lat/0010041)}

A numerical study of Goldstone-mode effects and scaling functions of the three-dimensional $O(2)$ model

J. Engels, S. Holtmann, T. Mendes and T. Schulze \dotfill \hbox{861 (hep-lat/0010004)}

Superconductivity with the meron-cluster algorithm

J.C. Osborn \dotfill \hbox{865 (hep-lat/0010097)}

\vskip 8mm
List of participants \dotfill \noeprint{869}

Author index \dotfill \noeprint{881}

General information \dotfill \noeprint{887}

\end{document}